\documentclass[aps,prl,reprint, amsmath, amssymb,superscriptaddress,nofootinbib]{revtex4-1}

\usepackage{bm}
\usepackage[retainorgcmds]{IEEEtrantools}
\usepackage{graphicx}
\usepackage{mathrsfs}
\usepackage{amsmath}
\usepackage{amssymb}
\usepackage{color}
\usepackage{amsfonts}
\usepackage{times,txfonts}
\usepackage{nicefrac}
\usepackage[colorlinks]{hyperref}

\usepackage[dvipsnames]{xcolor}
\definecolor{mypink}{rgb}{0.858, 0.188, 0.478}
\usepackage[authormarkup=none]{changes}
\definechangesauthor[name={Stella}, color=mypink]{s}
\definechangesauthor[name={Stefan}, color=purple]{n}
\definechangesauthor[name={Gabriel}, color=red]{g}
\definechangesauthor[name={Valerio}, color=green]{v}

\newcommand{\ud}{\,\mathrm{d}}

\DeclareMathOperator{\sech}{sech}
\DeclareMathOperator{\tr}{tr}

\newcommand{\F}{\mathcal{F}}

\newcommand{\bra}[1]{\ensuremath{\langle#1|}}
\newcommand{\ket}[1]{\ensuremath{\left|#1\right\rangle}}

\newcommand{\mean}[1]{\ensuremath{\left\langle #1 \right\rangle}}

\begin{document}

\title{Collisional quantum thermometry}
\date{\today}
\author{Stella Seah}
\affiliation{Centre for Quantum Technologies, National University of Singapore, 3 Science Drive 2, Singapore 117543.}
\affiliation{Department of Physics, National University of Singapore, 2 Science Drive 3, Singapore 117542.}
\author{Stefan Nimmrichter}
\affiliation{Max Planck Institute for the Science of Light, Staudtstra{\ss}e 2, 91058 Erlangen}
\author{Daniel Grimmer}
\affiliation{Institute for Quantum Computing, University of Waterloo, Waterloo, ON, N2L 3G1, Canada.}
\affiliation{Dept. Physics and Astronomy, University of Waterloo, Waterloo, ON, N2L 3G1, Canada.}
\author{Jader P. Santos}
\affiliation{Instituto de F\'isica da Universidade de S\~ao Paulo,  05314-970 S\~ao Paulo, Brazil.}
\author{Valerio Scarani}
\affiliation{Centre for Quantum Technologies, National University of Singapore, 3 Science Drive 2, Singapore 117543.}
\affiliation{Department of Physics, National University of Singapore, 2 Science Drive 3, Singapore 117542.}
\author{Gabriel T. Landi}
\email{gtlandi@if.usp.br}
\affiliation{Instituto de F\'isica da Universidade de S\~ao Paulo,  05314-970 S\~ao Paulo, Brazil.}

\begin{abstract}
We introduce a general framework for thermometry based on collisional models, where ancillas probe the temperature of the environment through an intermediary system. This allows for the generation of correlated ancillas even if they are initially independent. Using tools from parameter estimation theory, we show through a minimal qubit model that individual ancillas can already outperform the thermal Cramer-Rao bound. 
In addition, {\color{black}due to the steady-state nature of our model}, when measured collectively the ancillas {\color{black}always} exhibit superlinear scalings of the Fisher information. {\color{black}  This means that even collective measurements on pairs of ancillas will already lead to an advantage}. {\color{black}As we find in our qubit model, such a feature may be particularly valuable} for weak system-ancilla interactions.
Our approach sets forth the notion of metrology in a sequential interactions setting, and may inspire further advances in quantum thermometry.
\end{abstract}
\maketitle{}

Quantum metrology aims to employ resources such as entanglement \cite{Giovannetti2006}, coherence \cite{Pires2017} and squeezing \cite{Maccone2019}, to provide  improvements in the precision of a large variety of physical measurements  \cite{Pezze2016,Braun2017a,RoadMap2018}.
One such type, which stands out due to its universal importance, is thermometry \cite{Pasquale2018,Mehboudi2018}.
Several papers have discussed how quantum resources could be used to yield significant enhancements in temperature estimation \cite{Kiilerich2018,Jahromi2018,Campbell2017,Jevtic2015b,Mancino2017a,Correa2016,Hofer2017a,Hovhannisyan2018,Ivanov2019}. This could have a large impact particularly on low temperature applications such as ultra-cold atoms, trapped ions and superconducting devices, where  minimally-invasive thermometric methods are seldom available (c.f. the proposal in Ref.~\cite{Mehboudi2019} for a counterexample).

Quantum thermometry  can be cast under the general framework of parameter estimation theory. 
The goal is to estimate the temperature $T$ of a thermal environment by allowing it to interact with one or more ancillas, which are then subsequently measured. 
In the simplest measurement protocol, $N$ identical ancillas thermalize individually with the environment. The finite heat capacity $C$ of each ancilla will then limit the precision of the corresponding temperature estimate to a minimum uncertainty of $(\Delta T/T)^2 \geq k_B/ N C$, corresponding to the Cramer-Rao bound for thermal states \cite{Mehboudi2018,Pasquale2018}. 
The $1/N$ scaling of the temperature variance, also known as the standard limit, stems from statistically independent measurements of each ancilla.
In practice, the bound may not be attainable when the ancilla interacts strongly with its environment \cite{Mehboudi2019}.

In general, superlinear scalings in $1/N$ can be obtained with collective measurements on correlated ancillas, e.g.~by preparing an initially entangled $N$-ancilla state \cite{Giovannetti2006,Pires2017}. 
Alternatively, one may perform $N$ sequential measurements on the same ancilla \cite{Guta2011,Burgarth2015,DePasquale2017,Clark2018}, in such a way that the outcomes are correlated. 
Specifically in the context of thermometry, this method would generally not improve the accuracy of temperature estimation \cite{DePasquale2017}, but this could in principle  change using adaptive schemes \cite{Clark2018}.

In this Letter, we introduce an alternative framework of thermometry inspired by collisional models \cite{Scarani2002,Grimmer2016,Strasberg2018,DeChiara2018,Seah2019}, in which the quantum advantages arise from repeated interactions between a continuous stream of independently prepared ancillas and a system mediating the thermal contact with the environment.
We show that, at sufficiently high repetition rates and strong interactions, individual ancillas already surpass the thermal Cramer-Rao bound \cite{Jevtic2015b}. 
In addition, enhancements are also found by exploiting the quantum correlations that are created among multiple ancillas. The paper presents first the theoretical framework, then explicit results based on a minimal qubit model.    

\begin{figure}
\centering
\includegraphics[width=1.0\columnwidth]{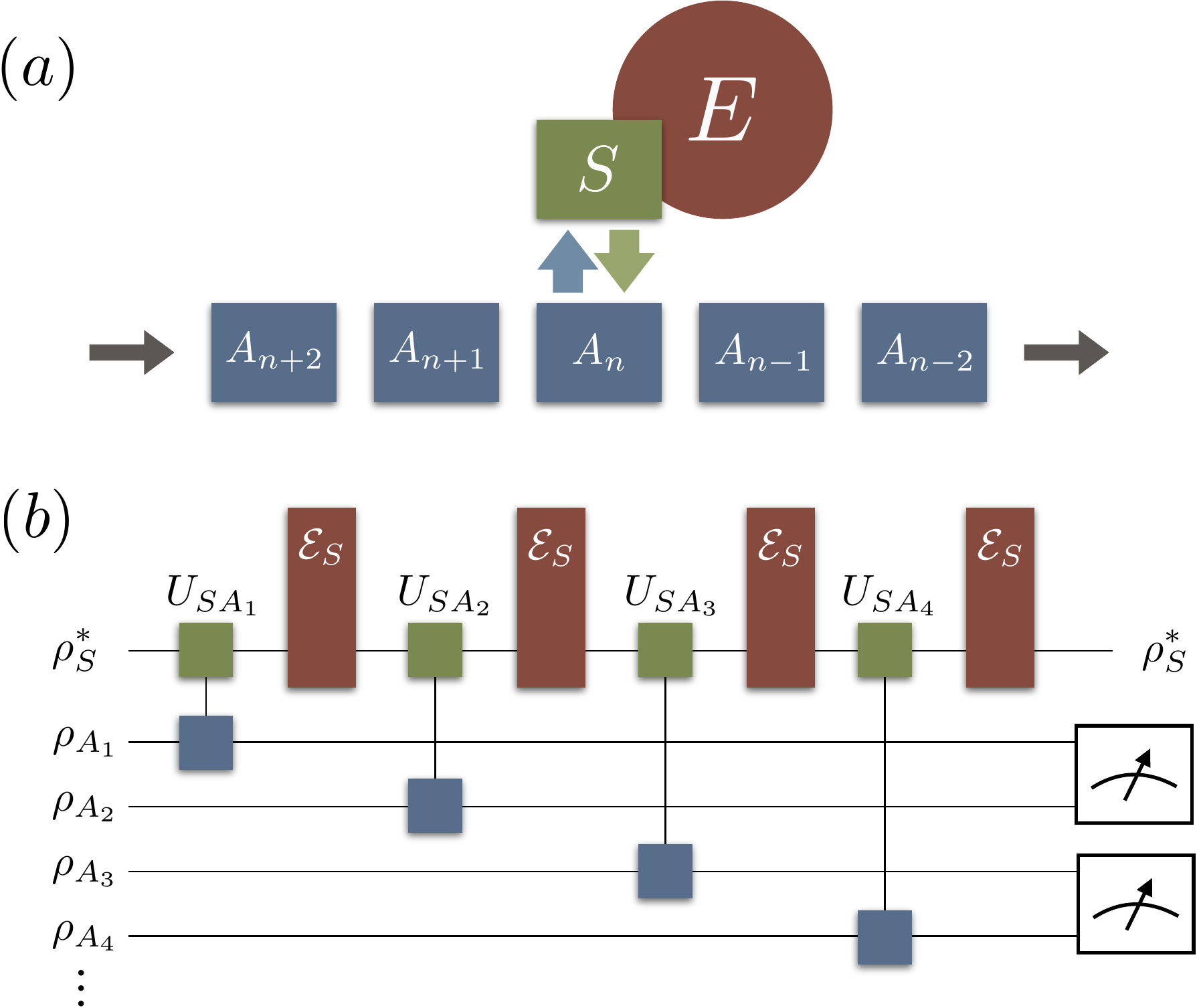}
\caption{\label{fig:diagram}
(a) Schematic diagram of collisional thermometry. A stream of ancillas sequentially interacts with a system $S$ which is coupled to a heat environment $E$ at temperature $T$.
(b) Exemplary circuit representation.
The system and the ancillas evolve stroboscopically in a sequence of $SE$ thermalization and pairwise $SA_i$ interaction steps.
The system is assumed to have converged to the fixed point $\rho_S^*$ of the stroboscopic map. 
Quantum advantages can be obtained by performing collective measurements on blocks of ancillas (illustrated here for blocks of length 2).
}
\end{figure} 

{\bf \emph{Formal framework} - }
We consider thermometry in a collisional model setting as  depicted in Fig.~\ref{fig:diagram}. A system $S$ is coupled to a thermal environment $E$ at temperature $T$ while interacting with a stream of independent and identically prepared (i.i.d.) ancillas $\{ A_1, A_2, \ldots\}$. Information about $T$ is encoded in the ancillas through $S$, which can then be retrieved by suitable measurements of a sufficiently large number of ancillas in order to build the statistics. 
Experiments with Rydberg atoms probing a thermally excited cavity, for instance, would be a natural physical implementation of this scheme \cite{Gleyzes2007}.

Assuming that the system-ancilla (SA) interaction time $\tau_{SA}$ is short compared to the system-environment (SE) coupling time $\tau_{SE}$ between subsequent ancillas and the characteristic thermalisation time, we can neglect the environment during the SA interaction. Hence, we describe the dynamics by alternating between the repeated application of a unitary map $\mathcal{U}_{SA_n} (\rho) = U_{SA_n} \rho U_{SA_n}^\dagger$ acting on the system and the $n$-th ancilla, followed by the thermal map $\mathcal{E}_S(\rho)$ acting only on the system during $\tau_{SE}$, see Fig.~\ref{fig:diagram}(b). Here, $U_{SA_n}=e^{-iV\tau_{SA}/\hbar}$ is the unitary operator generated by the SA interaction Hamiltonian $V_{SA_n}$.
The thermal map $\mathcal{E}_S(\rho)$ alone would drive the system towards the Gibbs state $\rho_S^\text{th} = e^{-H_S/k_B T}/Z$, where $H_S$ is the system Hamiltonian and $T$ is the environment temperature. 

The global state of $S, A_1, \ldots, A_n$ before the next $SA$ interaction is given by 
\begin{equation}\label{global_map}
\rho_{SA_1\ldots A_n} 
= \mathcal{E}_S \circ \; \mathcal{U}_{SA_n} \circ \ldots \circ \; \mathcal{E}_S \circ \; \mathcal{U}_{SA_1} \left(\rho_S \otimes \rho_{A}^{\otimes n} \right) ,
\end{equation}
where $\circ$ stands for map composition and $\rho_A$ is the initial state of the ancillas.
The reduced  state $\rho_S$ of the system therefore evolves stroboscopically according to the quantum channel
\begin{equation}\label{channel_system}
\rho_S(n) = \tr_{A_n}\big\{ \mathcal{E}_S \circ \mathcal{U}_{SA_n} \rho_S(n-1)\otimes \rho_A \big\} =: \Phi\big[ \rho_S(n-1)\big].
\end{equation}
The state $\rho_S(n)$ summarizes the information content about $E$ that the ancilla $A_{n+1}$ will have access to.

After sufficient $SA$ interactions, the system will no longer depend on its initial state and will eventually reach a fixed point $\rho_S^* = \Phi( \rho_S^*)$. This steady state contains information about $T$ that can be extracted by measuring subsequent ancillas. In general, $\rho_S^*$ will deviate from $\rho_S^{\rm th}$, which reflects the unavoidable measurement disturbance \cite{Seveso2018}.
In what follows, we will assume that the system has already reached  $\rho_S^*$, which eliminates transient effects and establishes translational invariance for subsequent states $\rho_{A_1\ldots A_N}$ of $N$ adjacent ancillas after they have interacted with $S$.

The key point in our approach is that even though the ancillas are initially independent, the final state $\rho_{A_1\ldots A_N}$ is generally correlated due to their common interaction with $S$ and the continuous exchange of information between $S$ and $E$. As a consequence, 
measurements on the collective state $\rho_{A_1\ldots A_N}$ of $N$ ancillas, e.g.~bipartite measurements ($N=2$) as in Fig.~\ref{fig:diagram}(b), could in principle give us more information about the temperature of the environment as compared to individual measurements of $N$ ancillas.

{\bf \emph{Temperature estimation } - } The quantum Fisher information (QFI) allows quantifying the precision of a thermometry scheme without reference to an explicit measurement protocol or experimental implementation \cite{Paris2009,Pezze2016,Braun2017a,Giovannetti2006}. Indeed, the standard deviation of any unbiased temperature estimator $\Delta T$ obeys the Quantum Cramer-Rao bound
\begin{equation}\label{QCR}
\Delta T^2 \geq \frac{1}{\mathcal{F}(T,\rho)},
\end{equation}
where $\mathcal{F}(T, \rho)$ is the QFI of temperature given the temperature-dependent quantum state $\rho$. 

Conventionally, $\rho$ is taken to be the state of a probe system that is in \emph{direct} thermal contact with the environment. 
{\color{black}Moreover, the inequality~(\ref{QCR}) is in general strict. The only exception is }
if the probe fully thermalizes. {\color{black}In this case} its information content about the temperature will be given by the thermal Fisher information (TFI)
\begin{equation}\label{TFI}
    \F(T,\rho) = \F_{\rm th} = \frac{C}{k_B T^2}, \quad C = \frac{\mean{H_A^2}-\mean{H_A}^2}{k_B T^2},
\end{equation}
with $H_A$ the ancilla Hamiltonian and $C$ its  heat capacity. 
{\color{black}In this special case, there exist estimators for which the inequality~(\ref{QCR}) is saturated in a single shot \cite{Mehboudi2018,Pasquale2018}.}
Repeating this process with $N$ probes then yields the aforementioned $1/N$-scaling of $(\Delta T/T)^2$.

Here, by contrast, $N$ ancillas acquire information about $E$ \emph{indirectly} and get correlated with each other through successive interactions with $S$. 
Information is extracted from the $N$-ancilla state $\rho_{A_1\ldots A_N}$ by means of a quantum measurement described by a positive operator valued measure (POVM) $\Pi_x$, where $x$ denotes the set of possible outcomes. 
This produces a probability distribution $p(x) = \tr\big\{\Pi_x \rho_{A_1\ldots A_N}\big\}$, from which one can construct the Fisher information  
\begin{equation}
F_N(\Pi, T,\rho_{A_1\ldots A_N} ) = \sum\limits_x p(x) \left( \frac{\partial}{\partial T} \ln p(x)\right)^2\,. 
\end{equation}
The QFI  is obtained by maximizing over all possible POVMs {\color{black}acting on a block of $N$ ancillas}, resulting in \cite{Paris2009}:
\begin{equation}\label{QFI}
\mathcal{F}_N(T,\rho_{A_1\ldots A_N}) = \max\limits_{\Pi} F_N(\Pi,T,\rho_{A_1\ldots A_N}) = \tr(\rho_{A_1\ldots A_N} \Lambda^2)
\end{equation}
where $\Lambda$ is the symmetric logarithmic derivative, which is a solution of the Lyapunov equation $\Lambda \rho_{A_1\ldots A_N} + \rho_{A_1\ldots A_N} \Lambda = 2 \partial_T \rho_{A_1\ldots A_N}$.

{\color{black}In Eq.~(\ref{QFI}), the length $N$ of the block of ancillas is arbitrary, with $\mathcal{F}_N$ representing the optimal precision that can be obtained from access to all possible measurement strategies on $N$ ancillas. 
For instance, $\mathcal{F}_1$ represents the best possible precision that can be obtained from information on individual ancillas, as encoded in  the completely marginalized product state $(\rho_A')^{\otimes N}$ 
(irrespective of whether the ancillas are correlated or not). 
In this scenario  the Fisher information for $N$ ancillas is simply  $N \mathcal{F}_1$ and will thus be linear in $N$ by construction. 
However, in general $\rho_A'$ will not be the thermal state of the environment, so that, as we show below, one can still obtain advantages over the TFI~(\ref{TFI}).
}

{\color{black}Another key advantage of our framework is that, once the system is operating in the steady-state, collective measurements on blocks of $N$ ancillas \emph{always} lead to advantages when compared with individual measurements. This is due to the fact that the steady-state is translation invariant, so that all ancillas individually contain the same amount of information (and hence collective states can only contain more). 
Or, in symbols, $\mathcal{F}_N \geq N \mathcal{F}_1$, with equality  when the ancillas are not correlated to each other. 
Collective measurements hence always lead to superlinear scalings. 
}


{\bf \emph{Qubit model } - } 
We now introduce a minimal qubit model that fits the described framework.
We take the system and ancillas to be resonant qubits with frequency $\Omega$, so that $H_S = \hbar\Omega \sigma_z^S/2$ and $H_A = \hbar\Omega \sigma_z^A/2$, where $\sigma_i$ are the Pauli matrices.
The $SE$ interaction is modeled by a standard quantum master equation which, in the interaction picture, reads as
\begin{equation}\label{MEq}
\frac{\ud \rho}{\ud t} = \mathcal{L}_S(\rho) =  \gamma (\bar{n}+1) \; \mathcal{D}[\sigma_-^S] + \gamma \bar{n}  \; \mathcal{D}[\sigma_+^S].
\end{equation}
Here $\mathcal{D}[L] = L \rho L^\dagger - \frac{1}{2} \{ L^\dagger L, \rho\}$, $\gamma$ is a temperature-independent coupling constant, and $\bar{n} = (e^{\hbar\Omega/k_BT}-1)^{-1}$. 
The thermal map $\mathcal{E}_S$ in Eq.~(\ref{global_map}) is obtained by integrating Eq.~(\ref{MEq}) over a certain time $\tau_{SE}$, yielding  $\mathcal{E}_S(\rho) = e^{\mathcal{L}_S \tau_{SE}} (\rho)$. 

Motivated by the framework of thermal resource theory \cite{Brandao2013}, we describe the $SA$ interactions by a partial SWAP Hamiltonian $V_{SA_n} =\hbar g (\sigma_+^S \sigma_-^{A_n} + \sigma_-^S \sigma_+^{A_n})$, which describes energy exchange as a thermal operation that does not require external work when on resonance.
We are left with the effective $SE$ and $SA$ couplings, $\gamma\tau_{SE}$ and $g\tau_{SA}$, the temperature $T$, and the initial ancilla state $\rho_A$ as the free parameters of our model. 

\begin{figure}
\centering
\includegraphics[width=1.0\columnwidth]{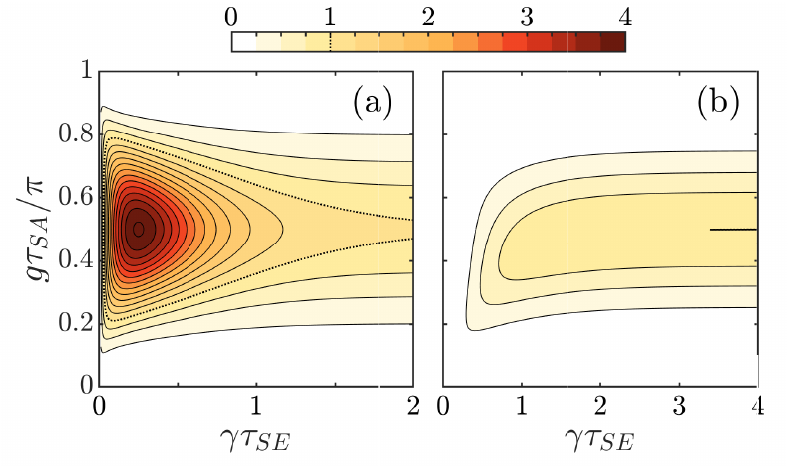}
\caption{\label{fig:F1}Quantum Fisher information $\F_1$ in units of  $\mathcal{F}_{\rm th}$($\approx 0.015$) for a single ancilla at steady state, as a function of the $SE$ coupling $\gamma\tau_{SE}$ and the $SA$ coupling $g\tau_{SA}$ for (a) ground-state ancillas and (b) $|+\rangle$-ancillas at $k_B T/\hbar\Omega =2$ (i.e.~$\bar{n}=1.514$). Dotted contour lines mark where $\F_1 = \F_{\rm th}$; coherent ancillas only exceed this in the limit of strong environment coupling and in close vicinity of a full swap.}
\end{figure}

{\bf \emph{{\color{black}Information acquired by a single ancilla}} - }
 {\color{black} We begin by addressing the scenario where one only has access to individual measurements on each ancilla. This means that irrespective of whether or not the ancillas are correlated, these correlations are assumed to be inaccessible to the experimenter.}
When the first ancilla interacts with the initially thermal state of the system, its QFI  will be bounded by the TFI \eqref{TFI}, which in this specific case reads $\F_{\rm th} = (\hbar \Omega/2 k_B T^2)^2 \sech^2\left(\hbar \Omega/k_B T\right)$.
For subsequent ancillas, this is no longer true since the system state does not remain thermal. 
{\color{black}Once the stroboscopic steady state $\rho_S^{*}$ is reached, however, the QFI for all ancillas become identical and equal to $\mathcal{F}_1$ in Eq.~(\ref{QFI}) (with the Fisher information for $N$ ancillas being simply $N \mathcal{F}_1$).}
Remarkably, as we now show, in this case each ancilla can actually gain \textit{more information} than $\F_{\rm th}$.

Figure \ref{fig:F1} depicts $\F_1/\F_{\rm th}$ as a function of the couplings $g\tau_{SA}$ and $\gamma\tau_{SE}$ for an exemplary temperature. In (a) and (b), we consider ancillas prepared in the ground state $\ket{g}$ and in the state $\ket{+} = (\ket{g}+\ket{e})/\sqrt{2}$, respectively. The TFI bound is always attained for full swaps ($g\tau_{SA}=\pi/2$) in the limit of perfect thermalization between subsequent ancillas ($\gamma\tau_{SE} \to \infty$), which is barely visible for the coherent case in (b). The greatest QFI values are generally attained also close to a full swap, but with ground-state ancillas. For instance, in (a) we see $\F_1/\F_{\rm th}\approx 4$ for $\gamma \tau_{SE} \lesssim 1$; the attainable ratio would grow with temperature. 

The reason for this enhancement is that, at steady-state operation and incomplete thermalization between adjacent interactions, the ancillas probe the environment temperature not only through the average system excitation (which saturates at $1/2$ for high $T$), but also through the effective thermal relaxation parameter $\Gamma = \gamma (2\bar{n}+1)\tau_{SE}$ (growing indefinitely with $T$). 
Explicitly for full swaps between $\rho_S^{*}$ and ground-state ancillas, the QFI reads as
\begin{equation}\label{single_anc_ratio}
\frac{\F_1}{\F_{\rm th}} = \frac{(\bar{n}+1)\left( e^\Gamma -1 +2\bar{n}\Gamma \right)^2}{e^{2\Gamma}(\bar{n}+1)-\bar{n}-e^\Gamma},    
\end{equation}
which makes evident the sensitivity on $\Gamma$.
A similar {\color{black}dependence on the relaxation rate}  was also observed in Ref.~\cite{Jevtic2015b}, but for a single qubit-environment interaction.

{\color{black} This effect is particularly advantageous at high temperatures ($k_B T \gg \hbar\Omega$), where Eq.~(\ref{single_anc_ratio}) exhibits a maximum at $\Gamma \approx 0.8$, in which $\F_1/\F_\text{th} \approx 0.65 (k_B T/\hbar\Omega)^2$. For $k_B T \gg \hbar\Omega$ one may therefore go \emph{significantly} above thermal sensitivity.
The existence of this maximum is a competition of two limiting cases.
}
In the Zeno limit,  $\Gamma \ll 1$, one gets  $\mathcal{F}_1\to 0$.
{\color{black}
Conversely, when $\Gamma \gg 1$ Eq.~(\ref{single_anc_ratio}) behaves as $\F_1/\F_{\rm th} \simeq 1 + 4 \Gamma \bar{n} e^{-\Gamma}$,  which tends to $1$, always from above.
This means that for large $\Gamma$ (which can always be obtained by choosing a sufficiently large $\tau_{SE}$) one will always be \emph{above} thermal sensitivity. 
}
For low temperatures, on the other hand, while it is possible to go above $\F_\text{th}$, no substantial improvements take place.

\begin{figure*}
\centering
\includegraphics[width=\textwidth]{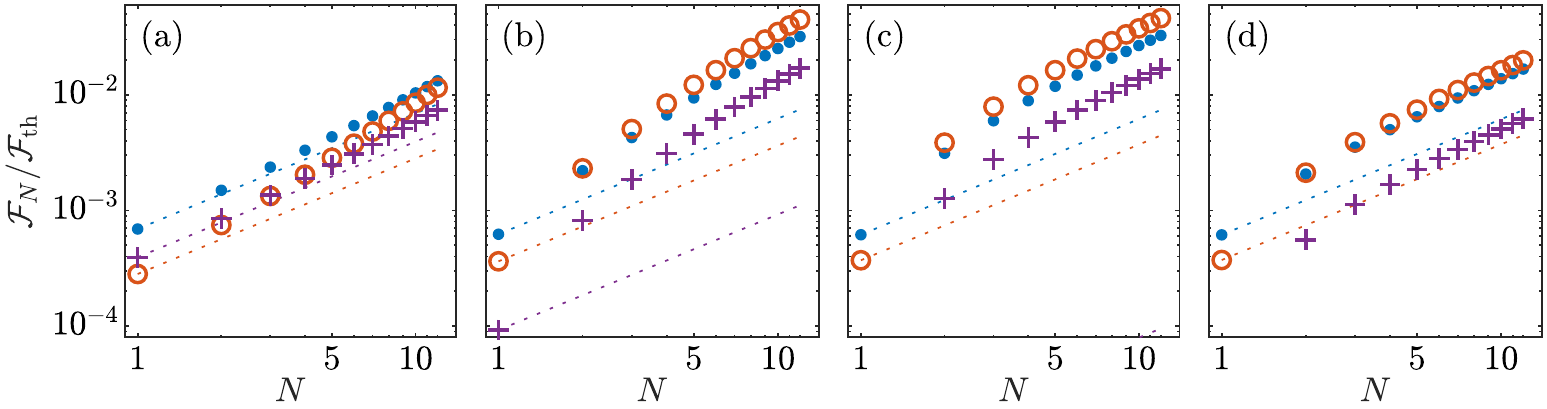}
\caption{\label{fig:FN} Log-log plot of quantum Fisher information $\mathcal{F}_N$ in units of $\mathcal{F}_{\rm th}$ ($\approx 0.015$) with the number of measured ancillas, $N$, for a fixed weak $SA$ coupling $g\tau_{SA}=\pi/100$ and $k_BT/\hbar\Omega = 2$. Panels (a) to (d) correspond to $SE$ couplings $\gamma\tau_{SE}=0.01$, 0.1, 0.4, and 1 respectively. Numerics is limited to 12 data points \cite{Sorensen2003,Footnote}. The markers (blue dots, red circles, and purple crosses) correspond to the use of $\ket{g}$, $\ket{e}$, and $\ket{+}$-state ancillas while the dotted lines give the {\color{black}individual measurement scenario}  $N\mathcal{F}_1$. Note that the purple line for $\ket{+}$-state ancillas cannot be seen in Panels (c) and (d) as $\mathcal{F}_1 < 10^{-5}\mathcal{F}_{\rm th}$ there.
}
\end{figure*}

{\bf \emph{{\color{black}Exploiting ancilla correlations} } - }
{\color{black}Substantial enhancements can also be obtained by exploiting the correlations between ancillas, as captured by the QFI $\mathcal{F}_N$ in Eq.~(\ref{QFI}) \cite{Footnote}.}
Figure \ref{fig:FN} shows {\color{black} the numerical behavior of} the QFI as a function of $N$ in the weak-coupling limit for various ancilla states (dots, circles, and crosses for $\ket{g}$, $\ket{e}$, and $\ket{+}$ respectively) and thermal couplings. 
{\color{black} As mentioned below Eq.~(\ref{QFI}), in our framework collective measurements always lead to an advantage in comparison with individual measurements. This is quite visible in Fig.~\ref{fig:FN}, where all choices of parameters always lead to a scaling that is larger than $N\F_1$ (dotted lines).}
Fig.~\ref{fig:FN}(a) represents the Zeno limit $\Gamma \to 0$, with $\gamma \tau_{SE} = 0.01$. In this case the superlinear enhancements in $\F_N$ are not substantial. Better results are achieved in the interval of intermediate {\color{black}$\gamma\tau_{SE}$, between 0.1 and 1 in (b)-(d). For greater $\gamma\tau_{SE} \gg 1$, correlations between successive ancillas vanish and $\F_N \to N\F_1$. }
We also observe that $\ket{e}$-ancillas are initially worse but eventually catch up to $\ket{g}$-ancillas around $N\sim 10$. Ancillas in $\ket{+}$-states exhibit the highest superlinear enhancement.

The superlinear improvement is caused by  quantum correlations that build up between subsequent ancillas. Indeed, a large improvement occurs between $\F_1$ and $\F_2$ when they become accessible.
The state of two adjacent ancillas, for instance, not only encodes information about $T$ in the individual excitations, but it also exhibits genuine distributed coherence \cite{Kraft2018}. 
For instance, if $\rho_A = \ket{g}\bra{g}$, we find the coherence 
\begin{eqnarray}\label{geeg_coherence}
 \bra{ge} \rho_{A_1A_2} \ket{eg} &=& e^{-\Gamma/2} \cos (g\tau_{SA}) p_A' ,
\end{eqnarray}
where
\begin{equation}
    p_A' = \bra{e} \rho_{A_{1}} \ket{e} =   \frac{\bar{n}(1-e^{-\Gamma}) \sin^2 (g\tau_{SA})}{(2\bar{n}+1)\left[1 - e^{-\Gamma}\cos^2 (g\tau_{SA})\right]},
\end{equation}
is the population of a single ancilla.
This excitation probability $p_A'$ is limited by the weak coupling angle $g\tau_{SA}$ and vanishes in the Zeno limit $\Gamma \to 0$. The coherence~(\ref{geeg_coherence}), however,  decays with an additional $e^{-\Gamma/2}$, so once again we should obtain {\color{black}optimal quantum enhancements at moderate $\Gamma$.}

{\color{black}
The enhancement  when going from $\mathcal{F}_1$ to $\mathcal{F}_2$ can be significant. To illustrate this, we consider the case 
 $g\tau_{SA} \ll 1$ and ground-state ancillas, for which we find
\begin{equation}
    \frac{\F_2}{2\F_{1}}  \approx 1+ \frac{(\bar{n}\Gamma)^2}{e^\Gamma - 1} + \mathcal{O} \left[ g\tau_{SA} \right]^2.
\end{equation}
This ratio has a maximum at $0.65 \bar{n}^2$, attained at $\Gamma \approx 1.6$.
For large $\bar{n}$ the enhancement can therefore be quite significant. 
Similarly, excited-state ancillas have their best enhancement of $0.65 (\bar{n}+1)^2$ at the same $\Gamma$-value.
The QFI $\mathcal{F}_N$ at this value of  $\Gamma$ is presented in Fig.~(\ref{fig:FN}(c). 
}

{\color{black} Exploiting the correlations between the ancillas may be particularly useful in the weak system-ancilla coupling regime ($g\tau_{SA} \ll 1$). 
In this case the process is much less invasive so that  $\mathcal{F}_1$ is generally quite small. 
It therefore becomes important to exploit the substantial enhancements stemming from correlations. }
For instance, in Fig.~\ref{fig:FN}(b), we get $\F_{10}/10\F_1 \approx 4$, 10, and 14 for $\ket{g}$, $\ket{e}$ and $\ket{+}$ respectively. 
{\color{black}We emphasize} that this does not require initial correlations  between the ancillas, which further demonstrates the virtue of the collisional thermometry scheme.

{\color{black}
{\bf \emph{Feasibility}}- The key assumption behind our setup is the dynamical nature of the sequential interactions. This does not mean, however, that one needs to have access to an infinite number of ancillas. 
A finite number together with the ability to recycle/reset them (e.g. by coupling them to a large bath) is  sufficient. 
For example, one can design the following protocol for $N=2$ ancillas:  interact them sequentially with the system, perform a collective measurement on both, and then reset them to their original state. Repeating this process over and over will generate precisely our collisional setup, but allowing one  to only exploit the advantages quantified by $\F_2$.

Finally, it is also worth mentioning that local measurements on the ancillas also lead to some correlations, although usually much smaller than those extractable from collective measurements. Notwithstanding, these correlations may be exploited in a similar way. 
}

{\bf \emph{Conclusions}}- We have introduced a framework for thermometry using collisional models. As an example, we have demonstrated that improved sensitivities of temperature can be reached even in a minimal qubit model. Specifically, if one is not constrained by short probing times or weak interactions, our scheme predicts a sensitivity that could outperform the thermal Cramer-Rao bound. In addition, our scheme also {\color{black}yields enhancements by exploiting the correlations developed between the ancillas.
In fact, due to the translational invariance of the stroboscopic steady-state, collective measurements will \emph{always} lead to an advantage when compared with individual measurements. }
Our framework opens the doors for metrology as a whole in the context of an open system, e.g.~measuring environment relaxation rates. A natural and interesting follow-up would be whether it is possible to push further these enhancements by considering probes with initial entanglement.

\acknowledgments{
\emph{Acknowledgments - }The authors acknowledge fruitful discussions with D. P. Pires, D. Soares-Pinto, A. Stokes, F. Binder, L. I. L. da Silva, D. Poletti, K. Amul, M. Genovese, A. Frota, K. Modi, K. Murch and S. Vinjanampathy.
This research is supported by the National Research Fund and the Ministry of Education, Singapore, under the Research Centres of Excellence programme. DG acknowledges support by  Natural Sciences and Engineering Research Council of Canada  (NSERC) through a Vanier Scholarship. GTL and DG acknowledge the Centre for Quantum Technologies in Singapore, where part of this work was developed, for both the financial support and  hospitality. }

\bibliography{addlit}

\end{document}